\documentclass[copyright,creativecommons]{eptcs}

\usepackage{latexsym}
\usepackage{amsmath}
\usepackage{amsfonts}
\usepackage{gastex}
\usepackage{color}

\newtheorem{theorem}{Theorem}[section]   
              
   \newtheorem{lemma}{Lemma}[section]      
    \newtheorem{corollary}{Corollary}[section]

\providecommand{\event}{GANDALF 2011} % Name of the event you are submitting to
\usepackage{breakurl}             % Not needed if you use pdflatex only.

\title{On $P$-transitive graphs and applications}
\author{Giacomo Lenzi
\institute{University of Salerno\\
Department of Mathematics\\
Fisciano (SA), Italy}
\email{gilenzi@unisa.it}
}
\def\titlerunning{On $P$-transitive graphs}
\def\authorrunning{G. Lenzi}
\begin{document}
\maketitle

\begin{abstract}
We introduce a new class of graphs which we call $P$-transitive graphs, lying between transitive and $3$-transitive graphs. First
 we show  that the analogue of de Jongh-Sambin Theorem is false for wellfounded $P$-transitive graphs; then we show that the $\mu$-calculus fixpoint hierarchy is infinite for $P$-transitive graphs. Both results contrast with the case of transitive graphs. We give also an undecidability result for an enriched $\mu$-calculus on $P$-transitive graphs. Finally, we consider a polynomial time reduction from the model checking problem on arbitrary graphs to the model checking problem on $P$-transitive graphs. All these results carry over to $3$-transitive graphs.
\end{abstract}

\section{Introduction}

The modal $\mu$-calculus, introduced in \cite{K83}, is a powerful logic for reasoning on systems, modeled as graphs. It is extensively used in computer science for verification of systems, both software and hardware. Formally, the $\mu$-calculus  is the extension of modal logic with the least fixpoint operator, $\mu$, and the greatest fixpoint operator, $\nu$. Intuitively, $\mu$ corresponds to inductive definitions, and is used to express liveness properties; and $\nu$ corresponds to coinductive definitions, and is used to express safety properties. Moreover, one is allowed to combine  $\mu$ and $\nu$, and this gives more power in general, e.g. we can express fairness properties with a $\nu$ followed by a $\mu$. When several fixpoint alternations are present in a $\mu$-calculus formula, the meaning of the formula is often hard to understand. However, 
it has been shown that the number of fixpoint alternations gives an infinite hierarchy on the class of all  graphs, see \cite{B96} and \cite{A99}. Things may change if one restricts attention to special classes of graphs.

One of the most important classes of graphs is the class of all wellfounded transitive graphs. The corresponding modal logic, G\"odel-L\"ob logic $GL$, is in fact the modal logic of provability in Peano Arithmetic, a logic deeply studied since the Seventies at least. Actually, it is known that in $GL$, fixpoints are redundant; this follows from the famous de Jongh-Sambin Theorem, see \cite{S85},   saying that $GL$  modal fixpoint equations, in a very general form, have a unique solution, and that this solution is expressed by a modal formula. So, in $GL$, the $\mu$-calculus collapses to modal logic. A recent simple proof of the de Jongh-Sambin Theorem has been given by Alberucci and Facchini, see \cite{AF}.

It is interesting to look for possible generalizations of the de Jongh-Sambin Theorem. It is known that there is a  collapse of the $\mu$-calculus on transitive graphs,  not to modal logic, but to the second level of the fixpoint hierarchy, see \cite{AF09}, \cite{DL10} and \cite{DO09}; moreover,
 on arbitrary wellfounded graphs,  it is known  that one type of fixpoint is enough to express the whole $\mu$-calculus (but fixpoints cannot be eliminated). In the same vein, we could for instance relax the notion of transitivity in some way. 

A natural relaxation of transitivity is $k$-transitivity, where $k$ is a positive integer: a graph is 
$k$-transitive if every point reachable in finitely many steps is reachable in at most $k$ steps. We maintain that $k$-transitivity has some practical motivation in relation to (bounded) model checking. Let us expand on this point.

Recall that the model checking problem for a logic is: given a finite graph (which models a finite-state system) and a formula of the logic, decide whether the formula is true in the graph. This formalizes the idea of proving the correctness of a system. When the logic is the $\mu$-calculus or some fragments of it, as is often the case, one speaks of $\mu$-calculus model checking. The theoretical status of $\mu$-calculus model checking is open, in the sense that it is in $NP$ (hence also in $co-NP$ by complementation) but it is not known whether it is in $P$. However, given the practical importance of model checking, it is interesting to look for practical model checking algorithms. Two important  techniques in this sense are the Symbolic Model Checking, see \cite{M}, and the Bounded Model Checking, see \cite{B}. 

In Symbolic Model Checking, the states and the relations of a system are modeled by  Boolean functions, and these functions are typically encoded in Binary Decision Diagrams, which are often much more succinct than the truth-table representation of the function. 

In Bounded Model Checking, instead, one operates as follows. Given a system, first one fixes a positive integer $k$ 
and considers only those states of the system which are reachable in $k$ steps from the initial state.  This means that if the graph is $k$-transitive, the entire space of reachable states is considered; if there are points not reachable in $k$ steps, they are discarded or considered later on. This approach  is sensible because, for instance, real computer networks are often $k$-transitive for quite small $k$, say  $k\leq 20$. Next one reduces the model checking problem to a propositional satisfiability problem, and the key point is to make this reduction efficient. Finally,  one uses a SAT solving algorithm. It must be remarked that, despite the theoretical intractability of the SAT problem, very efficient SAT solvers exist and are being constantly  developed, see \cite{BEATCS}. Thanks to $k$-transitivity, Bounded Model Checking often solves problems which are intractable with the usual techniques of Symbolic Model Checking (however one must admit that the contrary is also true).

In this paper, the final goal is to obtain some properties of $k$-transitive graphs. 
The main technical tool used for our results is a new notion of relaxed transitivity which we call $P$-transitivity, lying between transitivity and $3$-transitivity. Here $P$ denotes a special atomic proposition which may be true or false on a vertex of a graph. In some respect, $P$-transitive graphs behave better than $k$-transitive graphs, for instance there is a natural notion of  $P$-transitive closure of a graph. 

The focus of this paper is  on the $\mu$-calculus and its fragments on $P$-transitive graphs;  we give decidability and undecidability results as well as expressiveness results. Most of our results extend to $k$-transitive graphs (in particular to $3$-transitive graphs).

First, we prove that the satisfiability problem of the usual $\mu$-calculus on $P$-transitive graphs is decidable. Then, by contrast, we show that the satisfiability problem on $P$-transitive graphs for an enriched $\mu$-calculus essentially taken from \cite{BP} is undecidable. The undecidability result carries over to $3$-transitive graphs. 

The expressiveness results  are the following. First, the analogue of de Jongh-Sambin Theorem is false for wellfounded $P$-transitive graphs: on these graphs, the $\mu$-calculus does not collapse to modal logic. Hence the same happens for $3$-transitive graphs, 
in contrast with wellfounded transitive graphs  (the case of $2$-transitive graphs is open). Second, we show the infinity of the fixpoint hierarchy over $P$-transitive graphs, hence over $3$-transitive graphs: this is  again in  contrast with transitive graphs, where the hierarchy collapses to alternation $2$ (and again we do not know what happens for $2$-transitive graphs).

We remark that wellfoundedness is a rather strong property of  systems, being equivalent to termination of all computation paths; so wellfounded
 $k$-transitive graphs seem to have a mostly theoretical interest (e.g. we may view them as generalizations of the deeply studied wellfounded transitive graphs), whereas arbitrary $k$-transitive graphs, possibly with loops, are sufficiently general to model realistic networks.

We conclude the paper with some considerations on the model checking problem on $P$-transitive graphs. Via a polynomial time reduction, we show that the $\mu$-calculus model checking problem for $P$-transitive or $3$-transitive graphs is as difficult as for arbitrary graphs; we argue that for transitive graphs, finding such a polynomial time reduction (if any) would be a major breakthrough.

\section{Syntax}

\subsection{Modal Logic}

In modal logic over a finite set $A$ of atoms and one relation $R$, we have the modalities 
$\langle\ \rangle$ (diamond) and $[\ ]$ (box). Sometimes we will have several relations $R_1,R_2,\ldots$, in which case disambiguating notations $\langle R_1\rangle$, $\langle R_2\rangle$, etc. will be used, and similarly for boxes. We specify syntax for one relation modal logic as follows (syntax for many relations is analogous).
 
$$\phi::=a\ |\ \neg a\ |\ \phi\vee\phi\ |\ \phi\wedge\phi\ |\ \langle \ \rangle\phi\ |\ [\ ]\phi,$$ 
where $a\in A$. 

The modal depth of a formula is $md(\phi)$, defined by:
$md(a)=md(\neg a)=0$; $md(\neg\phi)=md(\phi)$; $md(\phi\vee\phi')=md(\phi\wedge\phi')=max(md(\phi),md(\phi'))$; and 
$md(\langle\  \rangle\phi)=md([\  ]\phi)=md(\phi)+1$.

For the purposes of the next sections, we find it convenient to fix an element $P$ of $A$, together with its negation $\overline{P}$, and to define four modalities $\langle PP\rangle$, 
$\langle P\overline{P}\rangle$, $\langle \overline{P}P\rangle$, $\langle \overline{PP}\rangle$, whose meaning is: $\langle PP\rangle\phi= P\wedge\langle\ \rangle (P\wedge\phi)$, and similarly for the other three modalities. 
It is also convenient to write $\langle P-\rangle\psi$ for  $\langle PP\rangle\psi\vee\langle P\overline{P}\rangle\psi$, and similarly for $\langle\overline{P}-\rangle$, $\langle -P\rangle$ and $\langle -\overline{P}\rangle$. 
Every diamond defined in this way has a corresponding box.

\subsection{The $\mu$-calculus}

The $\mu$-calculus is the extension of modal logic with two operators $\mu X.\phi(X)$ and $\nu X.\phi(X)$. So the syntax is
$$\phi::=a\ |\ \neg a\ |\ X\ |\ \phi\vee\phi\ |\ \phi\wedge\phi\ |\ \langle \ \rangle\phi\ |\ [\ ]\phi\ |\ \mu X.\phi(X)\ |\ \nu X.\phi(X),$$
where $X$ ranges over a countable  set of variables.

Without loss of generality, we can assume that for every variable $X$ occurring in a formula $\phi$ there is only one subformula in $\phi$ of the form $\mu X.\psi$ or $\nu X.\psi$. 

If $X$ is a variable in a formula $\phi$, and $\psi$ is another formula, we can replace $X$ with $\psi$ everywhere in $\phi$, as long as $\psi$ does not contain free variables $Y$ such that some occurrence of $X$ is in the scope of $\nu Y$ or $\mu Y$ in $\phi$ (i.e. no capture can occur). The resulting formula is called a composition of $\phi$ and $\psi$.

\subsection{Fixpoint hierarchy}

The $\mu$-calculus fixpoint hierarchy is given by the classes of formulas $\Sigma_n,\Pi_n,\Delta_n$  defined recursively as follows.
\begin{itemize}
\item 
$\Pi_0=\Sigma_0$ is the class of formulas without fixpoints;
\item $\Pi_{n+1}$ is the closure of $\Sigma_n\cup\Pi_n$ with respect to composition and greatest fixpoints;
\item $\Sigma_{n+1}$ is the closure of $\Sigma_n\cup\Pi_n$ with respect to composition and least fixpoints;
\item $\Delta_n=\Sigma_n\cap\Pi_n$.
\end{itemize}

The alternation depth of a formula $\phi$, denoted by $ad(\phi)$, is the least $n$ such that $\phi\in\Delta_{n+1}$.

\section{Semantics}

\subsection{Graphs and Kripke semantics} 

A graph is a pair $G=(V,R)$ where $V$ is a set of vertices and $R$ is a relation on $V$. By means of graphs we can give a Kripke semantics for the $\mu$-calculus, extending the classical one for modal logic. We sketch the definition of the semantics, see e.g. \cite{BS} for more details. 

We start with a graph $G=(V,R)$ plus a valuation function $val$ from $A$ to the powerset of $V$.

The semantics of a formula $\phi$ in $(G,val)$ will be a set of vertices of the graph, defined inductively as follows.   

Atoms are interpreted by their valuations. Note that according to our stipulations, a vertex may verify several atoms, not necessarily just one. Boolean operators are interpreted  as usual. For modal operators, $\langle\ \rangle\phi$ means that some successor satisfies $\phi$, and $[\ ]\phi$ means that every successor satisfies $\phi$. 

For fixpoints, $\mu X.\phi(X)$ is the least solution of the fixpoint equation $X=\phi(X)$, and $\nu X.\phi(X)$ is the greatest solution of the equation, where $X$ ranges over the subsets of $V$.

For modal logic and the $\mu$-calculus over several relations, the semantics is analogous, but it is based on multigraphs $(V,R_1,R_2,\ldots)$, where $R_1,R_2,\ldots$ are relations on $V$.

We adopt the abbreviations $\phi\rightarrow\psi$ for implication and $\phi=\psi$ for equivalence between formulas. 

\subsection{Paths and trees}

Let $G=(V,R)$ be a graph. If $x,y$ are vertices of $G$, a finite path of length $n$ from $x$ to $y$ is a sequence $x=x_0Rx_1Rx_2\ldots Rx_n=y$. An infinite path is an infinite sequence $x_0Rx_1Rx_2\ldots$

Given $x,y\in V$, we say that $x$ reaches $y$ in $n$ steps if there is a path of length $n$ from $x$ to $y$. 

A graph  is  called a tree if there is a vertex $r$ such that for every vertex $v$, there is a unique path from $r$ to $v$. The vertex $r$ is unique and is called the root of the tree. An example is the set $\{0,1\}^*$ of all finite binary strings, where $xRy$ holds if and only if $x$ is a prefix of $y$ and the length of $y$ is the length of $x$ plus one. The root is the empty string.

A graph $G$ is called wellfounded if it has no infinite path. $G$ is called transitive  
if, whenever a vertex $x$ reaches a vertex $y$, we have $xRy$. Transitive wellfounded graphs form the G\"odel-L\"ob class $GL$.

\subsection{$k$-transitive graphs}

Let $k$ be a positive integer. A graph $G$ is called $k$-transitive if whenever a vertex $x$ reaches a vertex $y$ in $k+1$ steps, $x$ reaches $y$ in at most $k$ steps. Sometimes, the smallest $k$ such that $G$ is $k$-transitive is called the reachability diameter of $G$.

Note that usual transitivity coincides with $1$-transitivity, and that for every $k$, the property of being $k$-transitive is  definable in first order logic. However, we will see that there are several differences between the model theoretic properties of transitive  and $k$-transitive graphs, at least for $k\geq 3$ (the border case $k=2$ apparently needs to be worked out yet).

\section{$P$-transitive graphs}

In this paper we want to obtain some properties of $k$-transitive graphs. To this aim we find it convenient to introduce a new class of approximately transitive graphs, which we call $P$-transitive graphs. Let $G=(V,R,P,\overline{P},\ldots)$ be a graph, vertex colored with two complementary atoms $P$ and $\overline{P}$ (and possibly other atoms). 
We could replace the atom $\overline{P}$ with $\neg P$, but we prefer to consider $\overline{P}$ as an atom for notational reasons. 

We say that the $P$-color of a vertex is $P$, if the vertex satisfies $P$, and $\overline{P}$ otherwise. A graph $G$ is called $P$-transitive if for every two vertices $x,y$ of $G$, if 
$x$ reaches $y$ and $x,y$ have the same $P$-color, 
 then $xRy$.

\begin{lemma}
Every transitive graph is $P$-transitive. Every $P$-transitive graph is $3$-transitive.
\end{lemma}

{\em Proof:} the first statement is easy. For the second, it is enough to prove that in a $P$-transitive graph, if $x$ reaches $y$ in $4$ steps, then $x$ reaches $y$ in at most $3$ steps as well. So let $G$ be a $P$-transitive graph, and assume $x$ reaches $y$ in $4$ steps, let the corresponding path be $xRz_1Rz_2Rz_3Ry$. Among the five vertices there are three of the same color $P$ or $\overline{P}$. Suppose there are three $P$'s. Then if we link the first and the last $P$, we have a path from $x$ to $y$ of length $3$ or less. If there are three $\overline{P}$'s the reasoning is analogous.

\rightline{Q.E.D.}

\begin{lemma} $P$-transitive graphs can be defined by a formula of first order logic.
\end{lemma}

{\em Proof:} a graph is $P$-transitive if and only if the following weaker, first order condition holds: given two points $x,y$ with the same $P$-color, if there is  path from $x$ to $y$ of length  $2$ or $3$, then $xRy$.   In fact, assuming the condition we can show by induction on $n\geq 3$ that: for every two vertices $x,y$ with the same $P$-color, if there is a path of length $n$,  then $xRy$. 
The base case $n=3$ is immediate.
 For the inductive step from $m<n$ to $n$, with $n\geq 4$, consider $x,y$ of the same color and a path of length $n\geq 4$ from $x$ to $y$, and take the first $5$ points. By case analysis, between them there are at least two points with the same color at distance $2$ or $3$ in the path, and if we link them,  we get a  path from $x$ to $y$ with a length $m<n$, hence the induction hypothesis applies.

\rightline{Q.E.D.}

Note that for every graph $G$ colored with $P$ and $\overline{P}$, there is a unique smallest $P$-transitive graph containing $G$, which is obtained by linking all pairs of vertices $x,y\in G$ having  the same $P$-color and such that $x$  reaches $y$. This graph  will be called the $P$-transitive closure of  $G$, and will be denoted by $PTC(G)$. 

By contrast, note that it is not possible to define a unique $k$-transitive closure of a graph for $k>1$. Consider for instance a graph $G$ with vertices $1,2,3,4,5$ and relations $(1,2), (1,3), (2,4), (3,4), (4,5)$. Then we can add either $(2,5)$ or $(3,5)$ to make $G$ $2$-transitive, so there is no smallest $2$-transitive graph containing $G$.

By using the concept of $P$-transitive closure we can show the finite model property for $P$-transitive graphs:

\begin{lemma} If a $\mu$-calculus formula $\phi$ is true in some $P$-transitive graph, then it is true in some finite $P$-transitive graph.
\end{lemma}

{\em Proof:} suppose $\phi$ is true in a $P$-transitive graph $T$. We can suppose that all diamonds are of the form $\langle\ PP\rangle$ or  $\langle\ P\overline{P}\rangle$ or 
$\langle\ \overline{P}\overline{P}\rangle$ or $\langle\ \overline{P}{P}\rangle$, and similarly for boxes. Let $\phi^P$ be the result of replacing $\langle\ {P}{P}\rangle\psi$ with 
$$\langle\ {P}{P}\rangle\psi\vee\langle\ {P}-\rangle\langle\ \rangle^*\langle-P\rangle\psi$$ 
and similarly for $\langle\ \overline{P}\overline{P}\rangle\psi$, where as usual, the Kleene star
$\langle\ \rangle^*\alpha$ means $\mu X.\alpha\vee\langle\ \rangle X$.

Then $\phi^P$ is also true in $T$ and, by the finite model property of the $\mu$-calculus, it is true in some finite graph $F$ and also in $PTC(F)$. Since $PTC(F)$ is $P$-transitive, $PTC(F)$ verifies $\phi$.

\rightline{Q.E.D.}

\begin{corollary} \label{cor:decid} The $\mu$-calculus is decidable on $P$-transitive graphs.
\end{corollary}

{\em Proof:} by the small model property of the $\mu$-calculus over arbitrary graphs, the graph $F$ in the previous proof can be taken with size at most exponential in $\phi$, and the size of $F$ and $PTC(F)$ is the same, so we have a small model property also for $P$-transitive graphs.

\rightline{Q.E.D.}

\section{An undecidability result}

In the previous section we have shown that the $\mu$-calculus is decidable on $P$-transitive graphs. Things change if one enriches the $\mu$-calculus with additional constructs. In this section we consider essentially the same enriched $\mu$-calculus considered in \cite{BP} and \cite{B+}, and we prove that it is undecidable on $P$-transitive graphs by following \cite{BP}. 

More precisely, we consider the $\mu$-calculus over one relation $R$, and we enrich it with the following operators:
\begin{itemize}
\item $N$ (standing for nominal), an atom whose interpretation in any graph must be a singleton;
\item the counting modalities $\langle\ \rangle^{>n}\phi$, meaning that there are more than $n$ successors verifying $\phi$, and  $[\ ]^{\leq n}\phi$, meaning that there are at most $n$ successors verifying not $\phi$;
\item the inverse counting modalities $\langle\ \rangle^{-,>n}\phi$, meaning that there are more than $n$ predecessors verifying $\phi$, and   $[\ ]^{-,\leq n}\phi$, meaning that there are at most $n$ predecessors not verifying $\phi$.
\end{itemize}

We added above some counting modalities to the $\mu$-calculus considered in \cite{BP},  because they seem natural (in fact they appear in \cite{B+})  and they are essential for our undecidability proof: in \cite{B+} it is shown that, if we take only the trivial case $n=0$ in the modalities above, the resulting $\mu$-calculus is decidable. 
On the other hand, in \cite{BP} the enriched $\mu$-calculus is shown to be undecidable assuming to have several functional relations, whereas here we want to use one single relation, and in order to fit the spirit of this paper, we also insist that this relation must be $P$-transitive.
So, in the same vein of \cite{BP}, our result is:

\begin{theorem}\label{thm:undec}  The enriched $\mu$-calculus over a single $P$-transitive relation is undecidable.
\end{theorem}

{\em Proof:} the idea is the same as \cite{BP}: we simulate domino systems. Here are some details.

A  domino system is a triple $D=(T,Hor,Vert)$, where $T$ is a finite set of tile types and $Hor,Vert\subseteq T^2$. Let ${\mathbb N}=\{1,2,3,\ldots\}$ be the set of all natural numbers. One says that $D$ paves the grid $ {\mathbb N}\times {\mathbb N}$ if there is a function (a paving) from ${\mathbb N}\times {\mathbb N}$ to $T$ such that any two consecutive tiles sharing a horizontal side satisfy $Hor$, and any two consecutive tiles sharing a vertical side satisfy $Vert$.

In analogy with \cite{BP}, given a domino system $D$, we construct effectively an assertion $\psi_D$ in the enriched $\mu$-calculus over a single relation, such that $D$ paves the grid if and only if $\psi_D$ is satisfiable by a $P$-transitive graph. Since it is undecidable whether a domino system paves the grid, the satisfiability problem for the enriched $\mu$-calculus on $P$-transitive graphs is also undecidable.

In \cite{BP} one considers the grid as a bigraph $B=({\mathbb N}\times {\mathbb N},l,v,N)$,
 where we have two relations $l$ and $v$  corresponding to the left and vertical edges  of the grid, and $N$ denotes the origin of the grid. A paving of the grid is represented by predicates $C_t\subseteq {\mathbb N}\times {\mathbb N}$ for each $t\in T$.

In our setting, the relations $l$ and $v$ are coded by a single $P$-transitive relation $R$ plus four atoms as follows. 
 We define the $\bf abcd$-grid as the grid plus a suitable vertex coloring with colors $\bf a,b,c,d$.
That is, we color the odd rows of the grid (starting from the first one) by $\bf a, b, a,b \ldots$ and the even rows by $\bf c,d,c,d,\ldots$. 
In particular, the nominal $N$ is decorated with $\bf a$. We let also $P=\bf a\cup d$ and $\overline{P}=\bf b\cup c$.

In \cite{BP} there is the construction of a $\mu$-calculus formula $\phi_D$, such that a bigraph $(V,l,v,C_t,N)$ satisfies $\phi_D$ if and only if it is isomorphic to a paving of the grid. Note that \cite{BP} introduces also a third relation $d$,  but $d$ can be defined away in the grid as $d=(l^-\circ (v^-\circ l^-)^*)\cap (B_v\times B_h)$, where $l^-$ and $v^-$ are the inverse relations, $\circ$ denotes concatenation, $*$ is the Kleene star and $B_h$ and $B_v$ are the horizontal and vertical axes of the grid (both axes are in turn $\mu$-calculus definable in the grid). 

Now in a graph $G=(V,{\bf a,b,c,d},R,C_t,N)$ we define the left relation $l_G=R\bf \cap((b\times a)\cup (a\times b)\cup (c\times d)\cup (d\times c))$, and the vertical relation $v_G=R\bf \cap ((a\times c)\cup (c\times a)\cup (b\times d)\cup(d\times b))$. This gives also an auxiliary relation $d_G$ via the definition above of $d$.

If we replace in $\phi_D$ the relations $l,v$ with $l_G,v_G$, we have a first $\mu$-calculus formula $\psi^1_D$ which holds on a graph 
$G=(V,{\bf a,b,c,d},R,C_t,N)$ if and only if $l_G,v_G,d_G$ and their inverses are functional, and the associated bigraph
$B_G=(V,l_G,v_G,C_t,N)$ satisfies $\phi_D$. In fact, each modality $\langle l_G\rangle\alpha$ is definable as $\bf a\wedge\langle\ \rangle (b\wedge\alpha)\vee b\wedge\langle\ \rangle (a\wedge\alpha)\vee c\wedge\langle\ \rangle (d\wedge\alpha)\vee d\wedge\langle\ \rangle (c\wedge\alpha)$, and similarly for $\langle v_G\rangle$ and the inverse modalities.

 Note  that, in order to  express that the relations $l_G,v_G$ and their inverses are functional, we need counting modalities. 
For instance, $l_G$ is functional if and only if we have the conditions
$\bf a\rightarrow [\ ]^{\rm \leq 1}b$,  $\bf b\rightarrow [\ ]^{\rm \leq 1}a$, $\bf c\rightarrow [\ ]^{\rm \leq 1}d$, $\bf d\rightarrow [\ ]^{\rm \leq 1}c$.  This is why we added counting modalities with respect to \cite{BP}.  Likewise, functionality of the auxiliary relation $d_G$ and its inverse follows from functionality of $l_G,v_G$ and their inverses.  

Also, note that $\psi^1_D$ depends only on the restriction of $R$ to $(P\times\overline{P})\cup(\overline{P}\times P)$.

Moreover, the sets $\bf a,b,c,d$ as defined in the $\bf abcd$-grid verify the following conditions:
\begin{itemize}
\item $\bf a,b,c,d$ are a partition of the grid;
\item  the origin is $\bf a$, formally $N\rightarrow \bf a$;
\item  every $\bf a$  has a  $\bf b$ father and a $\bf c$ son, formally $\bf a\rightarrow\langle\ \rangle^- b\wedge
\langle\ \rangle c$; 
\item  every node with a $\bf b$ son and a $\bf c$ father is $\bf a$, formally 
 $\bf \langle\ \rangle b\wedge
\langle\ \rangle^- c\rightarrow a$;
\item every $\bf b$ has an $\bf a$ father and a $\bf d$ son, and every node with a $\bf d$ father and an $\bf a$ son is $\bf b$;
\item  every $\bf c$ has a $\bf d$ father and an $\bf a$ son, and every node with an $\bf a$ father and a $\bf d$ son is $\bf c$;
\item  every $\bf d$ has a $\bf c$ father and a $\bf b$ son, and every node with a $\bf b$ father and a $\bf c$ son is $\bf d$.
\end{itemize}
Note that all conditions above are modal; let $\psi^2$ be the conjunction of all these conditions, and let $\psi_D=\psi^1_D\wedge\psi^2$.

Now suppose $\psi_D$ is satisfiable by a $P$-transitive graph. Let $G=(V,{\bf a,b,c,d},R,C_t,N)$ be a $P$-transitive graph verifying $\psi_D$.  Then the bigraph $(V,l_G,v_G,C_t,N)$ satisfies $\phi_D$ and is isomorphic to a paving of the grid.

Conversely, take a paving of the grid in $D$. Then there is a bigraph $B=(V,{\bf a,b,c,d},l,v,C_t,N)$ isomorphic to a paving of the $\bf abcd$-grid. Define $R\subseteq ({\mathbb N}\times {\mathbb N})^2$   by $R=PTC(l\cup v)$; then the graph $G=(V,{\bf a,b,c,d},R,C_t,N)$ satisfies
 $\psi^2$.  Now, we have $l=l_G$ and $v=v_G$. So the bigraph $B=(V,l_G,v_G,C_t,N)=B_G$ satisfies $\phi_D$, and $G$  verifies $\psi^1_D$ as well, hence $\psi_D=\psi^1_D\wedge \psi_2$ is satisfiable (by a $P$-transitive relation). 

\rightline{Q.E.D.}

\begin{corollary} The enriched $\mu$-calculus over a single $3$-transitive relation is undecidable.
\end{corollary}

{\em Proof:} if $\psi_D$ is satisfied by any relation $R$,  then it is satisfied by the $3$-transitive relation  $$R'=PTC(R\cap((P\times\overline{P})\cup( \overline{P}\times P))).$$

\rightline{Q.E.D.}

\section{A lower bound for wellfounded $P$-transitive graphs}

In this and the following sections we give some expressiveness results for the $\mu$-calculus on $P$-transitive graphs and $3$-transitive graphs. We begin with a result which demonstrates that the de Jongh-Sambin Theorem cannot be extended to our context.

\begin{theorem} The $\mu$-calculus does not collapse to modal logic over wellfounded $P$-transitive graphs.
\end{theorem}

{\em Proof:} let $WPT$ be the class of wellfounded $P$-transitive graphs. Denote by $\rightarrow_{WPT}$ and $=_{WPT}$ implication and equivalence in $WPT$.
Let $\phi^+$ be a $\mu$-calculus formula saying that there is an alternating $P\overline{P}$-path to some point verifying an atom $Q$, beginning with $P$. We can take
$\phi^+=P\wedge\mu X.Q\vee \langle P\overline{P}\rangle X\vee\langle\overline{P}P\rangle X$.
Let $\phi^-$ be the same property as $\phi^+$ but beginning with $\overline{P}$. We note the equivalences (over arbitrary graphs)
$$\phi^+=P\wedge (Q\vee \langle\ \rangle\phi^-);$$
$$\phi^-=\overline{P}\wedge (Q\vee \langle\ \rangle\phi^+).$$

 Suppose for an absurdity that $\phi^+$ is modal on wellfounded $P$-transitive graphs. Let $\alpha$ be a modal formula of smallest modal depth such that $\alpha=_{WPT}\phi^+$.
The following lemma helps us in simplifying $\alpha$. 

\begin{lemma}\label{lemma:simplify} Let $ \phi^+\rightarrow_{WPT} [\ ]\gamma_0\vee\delta$, where 
$$\delta=
[\ ]\delta_1\vee\ldots\vee [\ ]\delta_n\vee\langle\ \rangle\zeta_1\vee\ldots\vee \langle\ \rangle\zeta_m\vee\sigma,$$ 
and where $\sigma$ is a disjunction of atoms and negated atoms. Then either $P$ implies $ [\ ]\gamma_0\vee\delta$  in $WPT$, or $ \phi^+$ implies $\delta$ in $WPT$.
\end{lemma}

{\em Proof:} suppose that $P$ does not imply $ [\ ]\gamma_0\vee\delta$  in $WPT$.
Let $M$ be a $WPT$ model of $ \phi^+$ and $N$ be a $WPT$ model of 
$P\wedge\neg( [\ ]\gamma_0\vee\delta)$. We can suppose that  $M$ and $N$ have the same colors at the root. Let $M'$ be the disjoint union of $M$ and $N$ up to identifying the roots. Then $M'$ is $WPT$ and still verifies $ \phi^+$, so it verifies  $[\ ]\gamma_0\vee\delta$. Moreover $M'$ cannot verify $[\ ]\gamma_0$ because it contains $N$, and likewise it cannot verify any box of $\delta$, so $M'$ verifies some diamond of $\delta$; but $N$ does not satisfy any diamond of $\delta$, hence by exclusion $M$ satisfies some diamond of $\delta$, and we conclude that $M$ verifies $\delta$, as desired.

\rightline{Q. E. D.}

Suppose that $\alpha$ is in conjunctive normal form (CNF). 
By applying several times Lemma \ref{lemma:simplify}, we can rewrite 
 $\alpha$ as a positive Boolean combination of atoms and diamonds. In fact, since $\alpha$ implies $P$, we can suppose that one conjunct of the CNF is $P$. If $ [\ ]\gamma_0\vee\delta$  occurs in the CNF, then we replace it with $P$, if the first case of the lemma applies, and with $\delta$ if the second case applies. The formula we obtain is still equivalent to $\alpha$. Iterating the procedure, we eliminate all boxes from the CNF. 

Now the next lemma helps us to eliminate some conjunctions from $\alpha$.

\begin{lemma} \label{lemma:conj} Suppose that a conjunction $$\gamma=\sigma\wedge \langle\ \rangle\gamma_1\wedge\ldots\wedge\langle\ \rangle\gamma_n$$ is satisfiable in $WPT$ and implies $\phi^+$ in $WPT$, where $\sigma$ is a  conjunction of atoms and negated atoms. 
Then  $\sigma$ contains $P$. Moreover  either  $\sigma$ contains $Q$,  or 
$P\wedge\langle\ \rangle\gamma_i$ implies $\phi^+$ in $WPT$ for some $i$. 
\end{lemma}

{\em Proof:} since $\phi^+$ implies $P$, $\gamma$ implies $P$ in $WPT$. Hence, $\sigma$ contains $P$, otherwise we could take a $WPT$ model of $\gamma$ and  put $\overline{P}$ at its root, and the resulting model would  be   a $WPT$ model of $\gamma$ with $\overline{P}$ at the root, contrary to the fact that $\gamma$ implies $P$ in $WPT$.

Suppose $\sigma$ does not contain $Q$.   
Suppose for an absurdity that there is a model $M_i$ in WPT which verifies  $P\wedge\langle\ \rangle\gamma_i$ but not $\phi^+$, for each $i=1,\ldots,n$. Since $Q$ does not occur in $\sigma$, we can suppose that all $M_i$ do not verify $Q$,  have the same colors at the root, and these colors verify $\sigma$.  Let us merge $M_1\ldots, M_n$ at the root. The resulting model is in $WPT$ and verifies $\gamma$ but not $\phi^+$,  a contradiction. So $P\wedge\langle\ \rangle\gamma_i$ implies $\phi^+$ in $WPT$ for some $i$. 

\rightline{Q.E.D.}

By the previous results we have a modal formula $\alpha$,  a positive Boolean combination of atoms and diamonds, equivalent to $\phi^+$ in $WPT$. Put $\alpha$ in disjunctive normal form. 
By applying several times Lemma \ref{lemma:conj}, we obtain a disjunctive normal form for $\alpha$ whose disjuncts have the form $P\wedge Q$ or $P\wedge\langle\ \rangle\gamma$. We note that a disjunct $P\wedge Q$ must be present, otherwise the one point model decorated with $PQ$ would verify $\phi^+$ but not $\alpha$. So,  by taking the disjunction, we can write
 $$\phi^+=_{WPT}P\wedge(Q\vee\langle\ \rangle\gamma).$$
We want to prove $\phi^-=_{WPT}\gamma$.

First we show that $\gamma$ implies $\phi^-$ in $WPT$. 
In fact, suppose for an absurdity that a $WPT$ model $M$ verifies $\gamma$ but not $\phi^-$.  
Take a fresh root decorated $P\wedge\neg Q$ and attach $M$ as a child. Let $N$ be the resulting model and let $N'=PTC(N)$ be its $P$-transitive closure. Then $N'$ verifies $P\wedge(Q\vee\langle\ \rangle\gamma)$, hence $N'$ verifies $\phi^+$; 
but the root of $N'$ has no $\overline{P}$ child except possibly for the root of $M$, so $N'$  verifies neither $P\wedge Q$ nor $\langle\ \rangle \phi^-$, hence $N'$ does not verify $\phi^+$, a contradiction. In particular, $\gamma$ implies $\overline{P}$ in $WPT$.

Now let us show that $\phi^-$ implies $\gamma$ in $WPT$.
Suppose for an absurdity that a $WPT$ model  $M$ verifies $\phi^-$ but not $\gamma$.
 As above, construct $N$ and $N'$. Then $N'$ verifies $P\wedge \langle\ \rangle\phi^-$ , hence $N'$ verifies $\phi^+$ and $P\wedge \langle \ \rangle\gamma$, but the root of $N'$ has no $\overline{P}$ child except for the root of $M$, so $N'$  does not verify 
$ \langle \ \rangle\gamma$, a contradiction.

Summing up, $\phi^-=_{WPT}\gamma$. Now we use a  lemma:
\begin{lemma} Let $\alpha,\beta$ be two  formulas containing an atom $P$ and its negation $\overline{P}$. Assume $\alpha,\beta$ are equivalent in $WPT$. Let $\alpha',\beta'$ obtained by swapping $P$ and $\overline{P}$ in $\alpha$ and $\beta$. Then $\alpha'$ and $\beta'$ are also equivalent in $WPT$.
\end{lemma}

{\em Proof:} let $M$ be a $WPT$ model of $\alpha'$. Let $M'$ be obtained from $M$ by swapping the valuations of $P$ and $\overline{P}$. Note that $M'$ is still $WPT$. Then $M'$ satisfies $\alpha$, hence $M'$ satisfies $\beta$ because $\alpha$ and $\beta$ are equivalent in $WPT$, and $M$ satisfies $\beta'$. So, $\alpha'$ implies $\beta'$ in $WPT$, and symmetrically, $\beta'$ implies $\alpha'$ in $WPT$, and the two formulas are equivalent in $WPT$.

\rightline{Q.E.D.}

Let $\overline{\gamma}$ be the result of swapping 
  $P$ and $\overline{P}$ in $\gamma$. By the previous lemma, we obtain $\overline{\gamma}=_{WPT}{\phi^+}$, but $\overline{\gamma}$ has modal depth lesser than the starting formula $\alpha$, contrary to the choice of $\alpha$. This proves the theorem.

\rightline{Q.E.D.}

\begin{corollary} The $\mu$-calculus does not collapse to modal logic over wellfounded $3$-transitive graphs.
\end{corollary}

\section{Parity games}

\subsection{Definition}

In the next section we will give a $\mu$-calculus hierarchy result for $P$-transitive graphs. In this section we prepare the result of the next section by recalling parity games.

%The formulas of $\mu$-calculus, especially in the higher levels of the fixpoint hierarchy, are often %difficult to understand. Parity games give an insight to the meaning of $\mu$-calculus, besides %being close to the $\mu$-calculus  from an algorithmic point of view.

Recall that a {\em parity game} is determined by a tuple $G=(V=V_c\cup V_d,v_0,E,\Omega)$ where $V$ is a countable set of vertices, $v_0\in V$ is the initial vertex, $V_c,V_d$ are two disjoint sets, $E$ is a binary relation on $V$, and $\Omega:V\rightarrow \{0,\ldots,n\}$ is a priority function. We choose to call the players $c$ and $d$  like \cite{A99} (where $c$ stands for conjunction and $d$ stands for disjunction).
The play works as follows. Players $c$ and $d$ move along the graph. On $V_d$, player $d$ moves; on $V_c$, player $c$ moves. If either has no move, the other wins. 
Otherwise, in an infinite play, $d$ wins if the greatest priority seen infinitely often is even, and $c$ wins otherwise. There is a $\mu$-calculus $\Sigma_n$ formula $W_{n}$, due to Walukiewicz, which expresses the fact that player $d$ has a winning strategy in the parity game associated to $G$. 

\subsection{Evaluation games}

Let $A$ be a finite set of atoms. Let $G$ be a countable graph vertex colored with $A$, let  $v_0$ be a vertex of $G$, and consider a $\mu$-calculus formula $\phi$. Following essentially \cite{AF09}, we define  a parity game $E(G,v_0,\phi)$ (the evaluation game), such that player $d$ wins the game if and only if $G,v_0$ verifies $\phi$. The difference is that we simplify the definition of priority of a formula.

The positions are pairs $(v,\psi)$ where $v$ is a vertex of the graph and $\psi$ is a subformula of $\phi$. The initial position is $(v_0,\phi)$. 
The $c$ positions are $(v,\psi\wedge\chi)$, $(v,[\ ]\psi)$, $(v,X)$ and $(v,\sigma X.\psi)$, as well as $(v,a)$ where $a\in A$  and $v$ verifies $a$ in $G$. 
The other positions are $d$ positions. 

There are arrows from $(v,\psi\wedge\chi)$ or $(v,\psi\vee\chi)$ to $(v,\psi)$ and $(v,\chi)$. There are arrows from $(v,[\ ]\psi)$ or $(v,\langle \ \rangle\psi)$ to $(w,\psi)$ for every  
successor $w$ of $v$. There are arrows from $(v,X)$ to $(v,\sigma X.\psi)$ and from $(v,\sigma X.\psi)$ to $(v,\psi)$. 

The priority of a position $(v,\psi)$ depends only on $\psi$. The priority of formulas which are not fixpoints is $0$. The priority of $\nu X.\chi$ is 
 $ad(\nu X.\chi)$, if this number is even, and $ad(\nu X.\chi)-1$ if it is odd;
 the priority of  $\mu X.\chi$ is  $ad(\mu X.\chi)$, if this number is odd, and $ad(\mu X.\chi)-1$ if it is even. 

\subsection{$P$-Parity games}\label{Waluk}

We find it convenient to introduce an ad hoc variant of parity games for $P$-transitive graphs, called $P$-parity games.
The definition is the same as for parity games, except that vertices are partitioned in $P$-states and $\overline{P}$-states, and 
 only moves from $P$ to $\overline{P}$ or from $\overline{P}$ to $P$ are allowed.
There is a $\mu$-calculus $\Sigma_n$ formula $P-W_{n}$, similar to the usual parity game formula of  Walukiewicz, which expresses the fact that player $d$ has a winning strategy in the $P$-parity game associated to a graph $G$.

\section{A hierarchy result}

\begin{theorem} The $\mu$-calculus fixpoint hierarchy is infinite on $P$-transitive graphs.
\end{theorem}

{\em Proof:} for every positive integer $n$ define a set of atoms  $$A_n=\{c,d\}\times\{0,\ldots,n\}\times\{P,\overline{P}\}.$$ 
Let $M_n$ the metric space of all complete binary trees labeled with $A_n$, that is, the functions from $\{0,1\}^*$ to $A_n$, where the root is labeled $P$ and $P,\overline{P}$ alternate along every path. Note that in this case, each vertex verifies exactly only one atom (this is not always the case in our semantics of the $\mu$-calculus).
The distance function in $M_n$ is defined as $d(T,T')=1/{2^k}$ if $T,T'$ coincide up to depth $k$ but not on higher depth, and $d(T,T')=0$ if they are equal.
A contraction of $M_n$  is defined as a map $f:M_n\rightarrow M_n$ such that, for some constant $c<1$, $$d(f(T),f(T'))\leq c\cdot d(T,T').$$

Let $T\in M_n$.   Let $\phi$ be a $\mu$-calculus formula over the alphabet $A_n$ of class $\Sigma_n$. We construct an incomplete binary tree $I=I(T,\phi)$ over $A_n$ such that $PTC(T)$ models $\phi$ if and only if $I(T,\phi)$ verifies $P-W_n$. The idea is to encode the evaluation game $E(PTC(T),\phi)$ in the tree $I$.

We define  partial binary trees $I(T,u,\psi)$, with alternating $P$-colors, where $u$ is a vertex of $T$ and $\psi$ is a subformula of $\phi$, according to the following rules.
For brevity in this definition we denote $I(T,u,\psi)$ by $I(u,\psi)$. 

If $Q\in A_n$, we let $I(u,Q)$  be a complete binary tree labeled  $(d,0)$  if $T,u$ verifies $Q$, and $(c,1)$  if $T,u$ does not verify $Q$. 
If $X$ is a variable we let $I(u,X)$  be a node labeled $(c,0)$ with a child where we attach the tree $I(u,\sigma X.\psi)$. If $\psi=\sigma X.\chi$, then $I(u,\psi)$ is a node labeled $(c,p(\sigma X.\chi))$ (here $p$ denotes priority) with a child where we attach the  tree $I(u,\chi)$.

If $\psi=\psi_1\vee\psi_2$, then  $I(u,\psi)$ is a node labeled $(d,0)$ with two children where we attach $I(u,\psi_1)$ and $I(u,\psi_2)$. 
If $\psi=\psi_1\wedge\psi_2$, then  $I(u,\psi)$ 
is a node labeled $(c,0)$ with two children where we attach $I(u,\psi_1)$ and $I(u,\psi_2)$. 

For modalities, we suppose that $\langle\ \rangle\alpha$ is replaced everywhere in $\phi$ with  
$\langle PP\rangle\alpha\vee \langle \overline{ PP}\rangle\vee\alpha\vee\langle P\overline{P}\rangle\alpha\vee\langle \overline{P}P\rangle\alpha$, so that it is enough to consider the four modalities above, and similarly we proceed with boxes. 

If $\psi=\langle P\overline{P}\rangle\chi$, or $\psi=\langle PP\rangle\chi$, and $T,u$ does not satisfy $P$, then 
we let $I(u,\psi)$  be a complete binary tree labeled  $(c,1)$ (in fact, in this case $\psi$ is trivially false). Likewise we proceed if $\psi=\langle \overline{P}{P}\rangle\chi$, or $\psi=\langle\overline{P}\overline{P}\rangle\chi$, and $T,u$ does not satisfy $\overline{P}$. 
Assume $\psi=\langle P\overline{P}\rangle\chi$, and $T,u$  satisfies $P$, or $\psi=\langle \overline{P}P\rangle\chi$, and $T,u$  satisfies $\overline{P}$; let $v,w$ be the children of $u$ in $T$;   then 
we let $I(u,\psi)$  be a node labeled $(d,0)$ with two children $I(v,\chi)$ and 
$I(w,\chi)$. 

If $\psi=[P\overline{P}]\chi$, or $\psi=[PP]\chi$ and $T,u$ does not satisfy $P$, then 
we let $I(u,\psi)$  be a complete binary tree labeled  $(d,0)$  (in fact, in this case $\psi$ is trivially true). Likewise we proceed if $\psi=[ \overline{P}P]\chi$, or $\psi=[\overline{P}\overline{P}]\chi$, and $T,u$ does not satisfy $\overline{P}$. 
Assume $\psi=[ P\overline{P}]\chi$, and $T,u$  satisfies $P$, or $\psi=[\overline{ P}P]\chi$, and $T,u$  satisfies $\overline{P}$; let $v,w$ be the children of $u$ in $T$;   then 
we let $I(u,\psi)$  be a node labeled $(c,0)$ with two children $I(v,\chi)$ and 
$I(w,\chi)$. 

{\em  The most delicate point are the rules  for the modalities $\langle PP\rangle$ in a $P$-point, 
$\langle \overline{PP}\rangle$ in a $\overline{P}$-point, and the corresponding boxes.}

 We must ensure that the construction is a contraction in the metric space $M_n$. The idea is to ``stretch'' the coding tree. To this aim we consider the set $E\subseteq\{0,1\}^*$ of all positions of even height in the binary tree. To each $e\in E$ we assign an integer $n(e)$ so that:
\begin{itemize}
\item $n(e)$ is greater than the height of $e$;
\item $e\not=e'$ implies $n(e)\not=n(e')$.
\end{itemize}

Assume  $\psi=\langle PP\rangle\chi$, and $T,u$  satisfies $P$. For $e\in E$, let  $u_e$ be the $P$-successor of $u$ with position $e$ in the subtree $T,u$.
  Then $I(u,\psi)$ is a node labeled $(d,0)$ with an an infinite alternating chain, where for every $e\in E$, the tree $I(u_e,\chi)$ is attached as a child at position $n(e)$. The vertices of the chain are labeled $(d,1)$ (so that $d$ is forced to exit the chain eventually, otherwise he loses).

An analogous rule applies when 
$\psi=\langle \overline{P}\overline{P}\rangle\chi$, and $T,u$  satisfies $\overline{P}$, 
where $\overline{P}$-successors are considered instead of $P$-successors.

Likewise, 
assume  $\psi=[ PP]\chi$, and $T,u$  satisfies $P$. Then 
$I(u,\psi)$ is a node labeled $(c,0)$ with an an infinite alternating chain, where for every $e\in E$, the tree $I(u_e,\chi)$ is attached as a child at position $n(e)$. The vertices of the chain are labeled $(c,0)$  (so that $c$ is forced to exit the chain eventually, otherwise he loses).

An analogous rule applies when 
$\psi=[\overline{P}\overline{P}]\chi$, and $T,u$  satisfies $\overline{P}$, 
where $\overline{P}$-successors are considered instead of $P$-successors.

\medskip

We let $I_\phi(T)=I(r,\phi)$, where $r$ is the root of $T$, and where $r$ is decorated with $P$.
In order to obtain an element of $M_n$, i.e. a complete binary tree, we transform  $I_\phi(T)$ to $G_\phi(T)$ as follows:
\begin{itemize}
\item if a node $v$ labeled $c$ has only one child, we attach to $v$ a second child with a complete binary tree labeled $(c,0)$ (so that $c$ loses if he follows the second child);
\item if a node $v$ labeled $d$ has only one child, we attach to $v$ a second child with a complete binary tree labeled 
$(d,1)$ (so that $d$ loses if he follows the second child).
\end{itemize}

Now, as usual we have a contraction property:
\begin{lemma}
$G_\phi$ is a contraction in the complete metric space $M_n$. 
\end{lemma}

{\em Proof:}  first we show that, if two rooted trees $(T,u)$ and $(T',u')$ are equal up to depth $k$, then $I(T,u,\psi)$ and $I(T',u',\psi)$ are equal up to depth $k+1$. This can be shown by induction on $k$. We consider only the case $\psi=\langle PP\rangle\chi$.

 Suppose $(T,u)$ and $(T',u')$ are equal up to depth $k$.  Let $v$ be a $P$-node
of $T$  at some distance $d_v\leq k$ from $u$. Let $v'$ be a node of $T'$, such that the position of $v'$ in the binary tree rooted in $u'$ is the same position of $v$ in the binary tree rooted at $u$. Then $(T,v)$ and $(T',v')$ are equal up to depth $k-d_v$. By inductive hypothesis, $I(T,v,\chi)$ and $I(T',v',\chi)$ are equal up to $k-d_v+1$.  Now the truncation of $I(T,u,\psi)$ to level $k+1$ is a chain where some of the trees
$I(T,v,\chi)$ are attached at distances $D_v>d_v$ from $u$; and the truncation of $I(T',u',\psi)$ to level $k+1$ is also a chain where the corresponding trees $I(T',v',\chi)$ are attached at distance $D_v$ from $u'$. 
 So, the truncations of $I(T,u,\psi)$ and $I(T',u',\psi)$ at levels $k+1$ are equal.

In particular, if two trees $T,T'$ are equal up to depth $k$, then $G_\phi(T),G_\phi(T')$ are equal up to depth $k+1$, and $G_\phi$ is a contraction of $M_n$ with constant $1/2$.

\rightline{Q.E.D.}

By the previous lemma, the usual argument 
of \cite{A99} applies. That is, suppose $\neg (P-W_n)$ is equivalent to a formula $\phi\in\Sigma_n$ on $P$-transitive graphs. Then for every element $T\in M_n$ we have:
$$PTC(T)\models\phi\leftrightarrow G_\phi(T)\models P-W_n,$$
and since $P$-parity games allow only moves from $P$ to $\overline{P}$ and conversely, we have 
$$PTC(T)\models\phi\leftrightarrow PTC(G_\phi(T))\models P-W_n.$$
Now by the Banach Fixpoint Theorem we can pick $T_0\in M_n$ such that $T_0=G_\phi(T_0)$, and we have:
$$PTC(T_0)\models\neg( P-W_n) \leftrightarrow PTC(T_0)\models P-W_n,$$ a contradiction.
Hence, $P-W_n$ cannot be $\Pi_n$ on $P$-transitive graphs, and the hierarchy is infinite.

\rightline{Q.E.D.}

\begin{corollary} The $\mu$-calculus fixpoint hierarchy is infinite on $3$-transitive graphs.
\end{corollary}

As a further corollary, we have that Theorem 4.57 of \cite{DO09} does not extend to finite $P$-transitive or finite $3$-transitive frames. In fact, let $ML^*$ be the modal logic where the modalities are those considered in Observation 4.26 of \cite{DO09}; that is, if $\phi$ is a formula and $p$ is a finite set of propositional formulas (i.e. formulas without modalities or fixpoints), then
the modality 
$\langle\ \rangle^*_p\phi$ means that there is an infinite path from the current point along which $\phi$ and  all propositional formulas in $p$ are true infinitely often (note that here $*$ does not quite denote the Kleene star, but something similar).

We have:

\begin{corollary} On finite $P$-transitive or $3$-transitive graphs, the bisimulation invariant fragment of\hfill\break Monadic Second Order Logic does not coincide with $ML^*$. 
\end{corollary}

{\em Proof:} $ML^*$ is a fragment of 
the $\mu$-calculus with alternation depth $2$, so by the previous corollary, $ML^*$ does not subsume the $\mu$-calculus, and a fortiori it does not subsume the bisimulation invariant fragment of MSO, over finite $P$-transitive or $3$-transitive graphs.

\rightline{Q.E.D.}

\section{On model checking $P$-transitive graphs}

One of the most important open problems in $\mu$-calculus is the complexity of the model checking problem: given a finite graph and a formula, decide whether the graph satisfies the formula. 
\cite{J98} gives a $UP$ upper bound, where $UP$  is the class of nondeterministic polynomial Turing machines with at most one accepting path on each input.
Moreover, \cite{JPZ} gives a subexponential algorithm. However, it is open whether the problem is solvable in polynomial time.

We note that there is a natural polynomial time reduction from the $\mu$-calculus  model checking problem on arbitrary finite graphs to the $\mu$-calculus model checking problem on finite $P$-transitive graphs. More precisely:

\begin{theorem}  Let $G$ be a graph and let $\phi$ be a $\mu$-calculus formula. 
There is a $P$-transitive graph $G'$, computable in polynomial time from $G$, and a formula $\phi'$, computable in polynomial time from $\phi$, such that  
 $G$ models $\phi$ if and only if $G'$ models $\phi'$. 
\end{theorem}

{\em Proof:}  Let $P$ be an atom not occurring in $\phi$. Call $\phi'$ the result of replacing every diamond $\langle\ \rangle\psi$  of $\phi$ with $\langle\ P\overline{P}\rangle\psi\vee\langle\ \overline{P}{P}\rangle\psi$, and similarly for boxes. 

The graph $G'$ is defined as follows. The vertices are the pairs $(v,P)$ and $(v,\overline{P})$, where $v$ is a vertex in $G$. We put an edge between $(v,P)$ and $(w,\overline{P})$, and one between $(v,\overline{P})$ and $(w,{P})$, if there is an edge between $v$ and $w$ in $G$.
Finally we put an edge between $(v,P)$ and $(w,P)$, and between $(v,\overline{P})$ and $(w,\overline{P})$  for every pair of vertices $v,w$ of $G$.

We note that, in order to verify $\phi'$ in $G'$, it is enough to consider an alternating model checking game for $\phi'$  in $G'$, alternating in the obvious sense, and a player wins such an alternating game in $G'$ if and only if the player wins the ordinary model checking game for $\phi$  in $G$. 
  So, $G$ models $\phi$ if and only if $G'$ models $\phi'$.

\rightline{Q.E.D}

\begin{corollary} The model checking problem for $\mu$-calculus on arbitrary graphs reduces in polynomial time to the model checking problem on $3$-transitive graphs.
\end{corollary}

We note that proving the theorem above with $G'$ {\em transitive}, rather than $P$-transitive, would be a major breakthrough. In fact,
on transitive graphs, every $\mu$-formula is equivalent to one with alternation depth $2$, so by the algorithm of \cite{EL86}, the model checking problem for a transitive graph $G$ and a formula $\phi$ is quadratic in $G$ (this is only true for every fixed formula, because we must have the time to translate a formula into another with alternation depth $2$, a translation which may take exponential time). So,
if  we could construct a transitive graph $G'$ in polynomial time from $G$ and a formula $\phi'$ in polynomial time from $\phi$  such that $G$ models $\phi$ if and only if $G'$ models $\phi'$, then 
the time complexity of the model checking problem would have an upper bound polynomial in $G$ with a degree independent from $\phi$ (again plus the time, possibly exponential in $\phi$, spent in  reducing $\phi$ to a formula of alternation depth $2$);
hence, we would be very close to proving the conjecture that the model checking problem is solvable in polynomial time.

However, as noted by a referee, we cannot expect that a collapse of $\mu$-calculus to formulas of low alternation depth is sufficient to have a  polynomial time model checking algorithm: an example in this sense is 
measured $\mu$-calculus, see \cite{BKV04}, where variables range over measures on a set rather than subsets of a set. Every $\mu$-calculus formula is equivalent to a measured $\mu$-calculus formula with  least fixpoints only (i.e. without greatest fixpoints), but this is not sufficient to have a polynomial time model checking algorithm for the $\mu$-calculus.

\section{Conclusion}

In this paper we have introduced a class of graphs, called $P$-transitive graphs, and we have begun the study of its model theoretic properties. A possible development could be in the style of \cite{DO09}. For instance, this last paper gives a lot of results on bisimulation invariant fragments of first order logic on various classes of transitive graphs, and one can ask to what extent the theory carries over to $P$-transitive graphs or $k$-transitive graphs.

We note that, like for transitive graphs, finite irreflexive $P$-transitive graphs coincide with 
finite wellfounded $P$-transitive graphs. For this class of graphs, probably
the theory of \cite{DO09} applies, and in particular Theorem 4.11 and its proof go through, just by replacing everywhere transitive closure with $P$-transitive closure (whereas, for $k$-transitive graphs, we have the problem of the non-uniqueness of the $k$-transitive closure). 

A challenging task in the same vein would be to find the bisimulation invariant fragment of first order logic for finite $P$-transitive graphs or $k$-transitive graphs. 

Note that for the class of all  (possibly infinite) $P$-transitive graphs, the bisimulation invariant fragment of first order logic is modal logic, because the class of $P$-transitive graphs is first order definable and van Benthem's Theorem applies, see \cite{DO09}. The same holds for $k$-transitive graphs.

We have seen that the extended modal logic $ML^*$ used in \cite{DO09} for finite transitive graphs is not sufficient to capture bisimulation invariant monadic second order logic on finite $P$-transitive graphs or $k$-transitive graphs. Probably the same happens for bisimulation invariant first order logic, because strongly connected components in these graphs are too complicated to be described by formulas of $ML^*$. However, one can look for other fragments of the $\mu$-calculus with bounded alternation depth which can replace $ML^*$ in this respect.

 Finally, an application of this paper to transitive graphs is expected: it seems plausible that the undecidability result in Theorem \ref{thm:undec} for $P$-transitive relations can be strenghtened to {\em transitive} relations, with a similar coding trick. In fact, the grid can be represented by a relation  where every vertex has only entering or only exiting edges: such a relation is vacuously transitive. However this result will be possibly presented elsewhere.

\bibliographystyle{eptcs} % or whatever you prefer

\end{document}